# Resolving structural changes and symmetry lowering in spinel FeCr$_2$S$_4$

Donald M. Evans[1], Ola G. Grendal[2], Lilian Prodan[1,3], Maximilian Winkler[1], Noah Winterhalter-Stocker[4], Philipp Gegenwart[4], Somnath Ghara[1], Joachim Deisenhofer[1], István Kézsmárki[1], and Vladimir Tsurkan[1,3]

[1]*Experimental Physics V, Center for Electronic Correlations and Magnetism, Institute of Physics, University of Augsburg, Universitaetsstrasse 1, 86159 Augsburg*
[2] *European Synchrotron Radiation Facility, 71 avenue des Martyrs, 38000 Grenoble, France*
[3]*Institute of Applied Physics, str. Academiei 5, MD 2028, Chisinau, R. Moldova*
[4]*Experimental Physics VI, Center for Electronic Correlations and Magnetism, Institute of Physics, University of Augsburg, Universitaetsstrasse 1, 86159 Augsburg*

(Dated 01.03.2022)

**Abstract:**
The cubic spinel FeCr$_2$S$_4$ has been receiving immense research interest because of its emergent phases and the interplay of spin, orbital and lattice degrees of freedom. Despite the intense research, several fundamental questions are yet to be answered, such as the refinement of the crystal structure in the different magnetic and orbital ordered phases. Here, using high-resolution synchrotron powder diffraction on stoichiometric crystals of FeCr$_2$S$_4$ we resolved the long sought-after cubic to tetragonal transition at ~65 K, reducing the lattice symmetry to *I4$_1$/amd*. With further lowering the temperature, at ~9 K, the crystal structure becomes polar, hence the compound becomes multiferroic. The elucidation of the lattice symmetry throughout different phases of FeCr$_2$S$_4$ provides a basis for the understanding this enigmatic system and also highlights the importance of structural deformation in correlated materials.

## I. Introduction

Over the last 50 years FeCr$_2$S$_4$, a prominent member of the large cubic spinel family, has been reported to exhibit a rich variety of physical phenomena emerging from the complex interplay of spin, orbital, and structural degrees of freedom. These fascinating phenomena include colossal magneto-resistance [1], half-metallicity [2], dynamic and static Jahn-Teller effects [3,4,5,6,7], Coulomb-enhanced spin-orbit interaction [8], topological anomalous Hall effect [9], multiferroicity [10,11], and magnetostructural transformations [10, 12,13,14]. However, this compound is highly sensitive to subtle changes in stoichiometry, which makes it difficult to establish the microscopic origins behind these phenomena. Specifically, the changes in stoichiometry cause notable alteration to the magnetic and structural properties. For instance, the phase transition at 9 K, recently documented in the magnetization and specific heat also in FeCr$_2$S$_4$ single crystals [14], is only observed in samples with stoichiometry close to ideal. Indeed, it can be fully suppressed by defects or site interchange of the cations of the order of 1 - 2 % [15,16,17,18, 19], which can rarely be excluded during the synthesis and crystal growth of spinel-type materials.

Despite the challenges of this intriguing system, magnetic properties in the low-field phases are believed to be well characterized. The stoichiometric FeCr$_2$S$_4$ is reported to order ferrimagnetically at $T_C$ ~ 165 K [14, 20]. Below $T_C$ down to $T_m$ ~ 65 K, the magnetic structure is collinear with an antiferromagnetic arrangement of spins of the tetrahedrally coordinated Fe$^{2+}$ and octahedrally coordinated Cr$^{3+}$ ions [10,21]. Below $T_m$ ~ 65 K, the collinear spin structure transforms into an incommensurate one, which remains the stable spin configuration down to the lowest temperature [22]. The temperature $T_m$ has previously been associated with local structural distortions attributed to dynamic Jahn-Teller (JT) correlations [20]. At $T_{OO}$ ~ 9 K, there was an unconventional first-order magnetostructural transition reported, which is associated with a peak in the specific heat and ~ 20 % reduction in the magnetic anisotropy [12,14,20].

Behind these well-established magnetic transitions, the low temperature space group symmetry has remained an enigma over the years. Some indications for the deviations from the cubic $Fd\bar{3}m$ symmetry, structural distortions, and structural transition at low temperatures have been provided by earlier Mössbauer measurements [4,5,6,7]. Additional hints for the structural anomaly at around $T_m$ came out from the ac susceptibility [23] and transmission-electron microscopy [24] studies, although previous x-ray and neutron powder-diffraction studies have not detected any structural anomaly below $T_{OO}$ anticipated for the static JT effect [21,25]. Even high-resolution synchrotron x-ray diffraction on stoichiometric polycrystals could not resolve the reduction of the symmetry with respect to the cubic structure, noticing however, the broadening of the diffractions peaks below $T_m$, which reaches a maximum at $T_{OO}$ [20]. Further insights have been provided by recent reports on the appearance of the multiferroic polar state below $T_{OO}$ [10,11] and on optical properties in the THz and FIR frequency range [13,26], which evidenced the reduction of the lattice symmetry below $T_m$ and $T_{OO}$, though the symmetry of the low-temperature phase remained unresolved.

In this work, based on the results of high-resolution synchrotron x-ray diffraction study performed on stoichiometric $FeCr_2S_4$ we establish that the low-temperature structure is described by a polar subgroup of the tetragonal $I4_1/amd$ space group. The symmetry lowering from the cubic $Fd\bar{3}m$ to the tetragonal $I4_1/amd$ is resolved below $T_m$ ~ 65 K. We correlate our structural data with the data of the thermal expansion, electronic, magnetic and heat capacity measurements. The pyrocurrent measurements allow us to conclude that below $T_{OO}$ the symmetry is further reduced to a polar subgroup of $I4_1/amd$. This holistic view of the phase transitions allows us to analyse the key traits for each transition, to offer an understanding of the crystal as a whole, crucial for future experimental and theoretical work.

## II. Experimental methods

All experiments presented here were performed on stoichiometric single crystals grown by the chemical transport reactions as described in Ref. [14].

**x-ray powder diffraction**

To prepare a sample for the diffraction study, single crystals of $FeCr_2S_4$ were selected, based on having high quality growth faces, and ground to a powder. The powder was annealed in evacuated quartz ampoule at 400 °C for two weeks, to remove any stress eventually induced by the grinding, before being stored and transported in vacuum conditions. The powder was filled into a 0.5 mm boron-silicate glass capillary. High-resolution powder x-ray diffraction (XRD) data were collected in transmission geometry on the high-resolution setup at the ID22 beamline, at European Synchrotron Radiation Facility (ESRF), Grenoble, France [27,28]. The wavelength was calibrated using silicon (NIST, 640c) to 0.35421748 Å (35 keV). Data were collected at 4, 7, 9, 15, 35, 60, 80, 100, 140, 165, and 190 K with the temperature being controlled with a liquid Helium cooled cryostat, and additional spectra were collected at ambient conditions (295 K) outside the cryostat. To reduce the effect of beam heating, the capillary was kept open at one end and exposed to the helium-flow atmosphere; in addition, the beam was attenuated for the experiments performed at 90 K and below. We note that the temperature reported is measured on the helium flow, and so the actual sample temperature is expected to be slightly higher than the stated value, due to heating from the beam.

Rietveld refinements and space group and lattice parameter determination of the low temperature phase were done using *TOPAS* (version 7). The most probable candidate structure for the low temperature phase was found by combing the result from the diffraction pattern indexing and group-subgroup relations from *Bilbao Crystallographic server* [29].

**Resistivity and polarization**

A single crystal with well-defined {111} facets was selected from the same batch used in the powder diffraction studies, polished down to thin slab parallel the (111) plane of the cubic setting. For both resistivity and pyrocurrent measurements, the electrical contacts were applied on these (111) faces with silver paint and gold wires. The electrical resistivity experiments were performed in the frame of conventional four-probe method in the temperature range from 10 K to 300 K using the resistivity option of the commercial Quantum Design physical property measurement system (PPMS). The pyrocurrent measurements were carried out in a capacitor-contact geometry using Oxford Helium flow cryostat together with the Keysight B2987A electrometer. This enables measurements between 300 and 2 K with the application of a magnetic field up to 14 T. The pyrocurrent data were obtained after cooling the sample from 20 K to 2 K in the presence of electric ($E_p$ = 1 kV/cm) and magnetic ($H_p$ = 4 T) poling fields. The poling fields were applied parallel to the crystallographic <111> direction of the cubic setting. The pyroelectric current was measured in constant magnetic fields after switching off the electric field and keeping the sample short circuited for at least 2 min while heating up with a constant rate of 4 K/min. The electric polarization was obtained by integration of the pyrocurrent with respect to time.

**Thermal expansion**

Thermal expansion was measured with the aid of a compact ultra-high-resolution capacitive dilatometer in the Quantum Design physical property measurement system (PPMS) [30]. A single crystal with well-defined {111} facets was oriented and polished to prepare a plane parallel sample with <100> orientation. For this measurement a sample with a thickness of 0.33 mm was used. The cell background was subtracted as it is described in [31]. For that purpose, a sample of pure copper with the same thickness as the sample was measured. To ensure good thermalization, the thermal expansion was measured from 2 to 200 K with a sweeping rate of 0.3 K/min. Several up and down sweeps were measured to verify the measured behavior.

**Magnetization and specific heat**

The magnetic properties have been studied on single-crystalline samples using a SQUID magnetometer (Quantum Design MPMS 3) in static fields applied parallel to the cubic <111> axis. The specific heat was measured in the commercial Quantum Design physical property measurement system (PPMS).

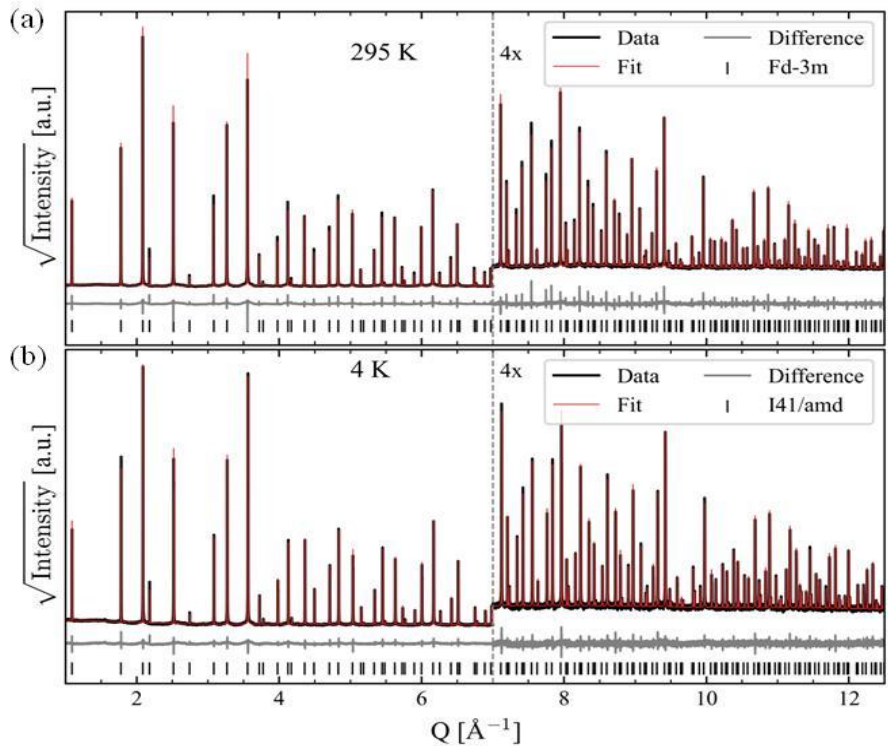

FIG 1: (a) Room temperature and (b) 4 K synchrotron high-resolution x-ray powder diffraction spectra of $FeCr_2S_4$.

### III. Experimental results

**x-ray powder diffraction**

We start by confirming the room temperature structure using the high-resolution diffraction data presented in Fig. 1(a). The recorded data (black lines) are fitted well (red lines) with the expected room temperature space group $Fd\bar{3}m$, no. 227, $a = 9.99816$ Å, $Z = 8$. The difference between fitted and expected values are plotted as grey lines and the expected peak positions are given by the black ticks. These data show no evidence for a secondary phase or other impurities. The data collected at base temperature, 4 K, are presented in Fig. 1(b) and the indexing show a good fit with the tetragonal space group $I4_1/amd$ (no. 141, $a = 7.0577$ Å, $c = 9.97793$ Å, $Z = 4$). Cubic, tetragonal, orthorhombic and hexagonal space groups were also searched for when indexing, however, no cubic or hexagonal space group were found to match all the observed peaks, and none of the orthorhombic space groups gave any significant improvement of the refinements over the proposed tetragonal space group.

Having successfully identified the high- and low-temperature structures, we collected additional spectra to find the structural transition temperature and see if any of the other known transitions affect the lattice parameters. This is best illustrated by the temperature dependency of the lattice parameters, presented in Fig. 2(a). Note, the same fitting approach was used at all temperatures. Fig. 2(a) shows a nearly linear thermal contraction of the lattice parameter $a$ (see fitted grey line) down to 80 - 60 K, below which the rate of thermal contractions changes. The grey dashed line in Fig. 2(a) terminates at low temperatures, where the fitting was changed from cubic to tetragonal structure, as discussed and justified below. Interestingly, the tetragonality (the difference between $a$ and $c$ lattice parameters) shows a monotonous increase until 7 K, below which it reduces.

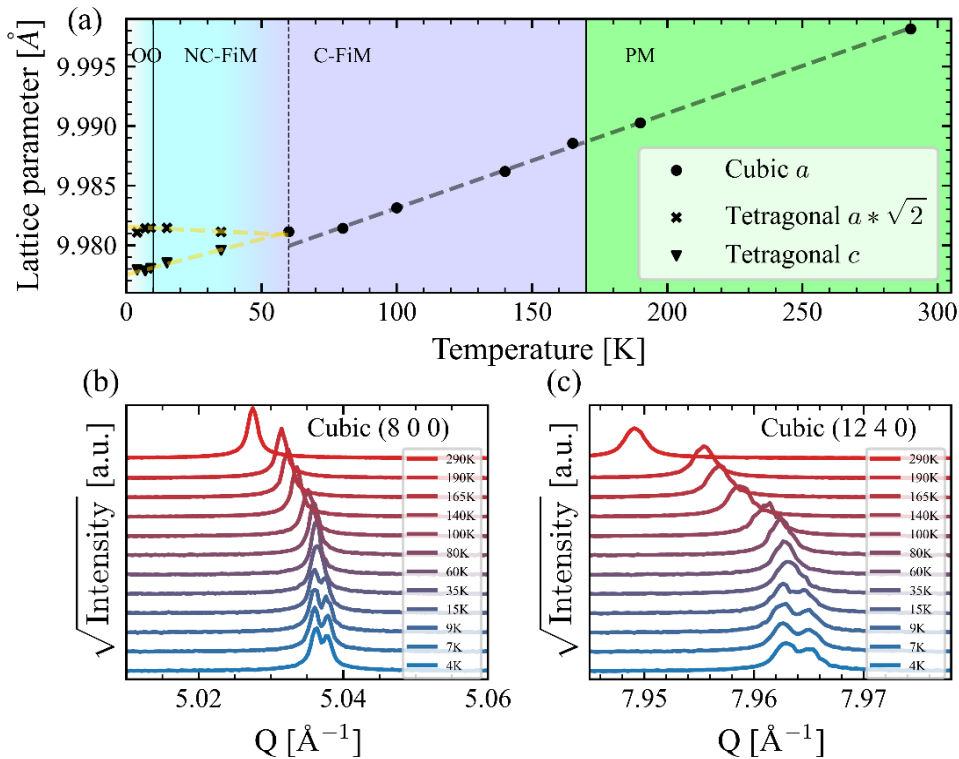

FIG 2: (a) Temperature dependence of the lattice parameters. Vertical grey dotted line shows separation between the use of the cubic and tetragonal space group for the Rietveld refinements. PM, C-FiM, NC-FiM, and OO mark respectively, cubic paramagnetic, pseudo-cubic collinear ferrimagnetic, tetragonal non-colinear ferrimagnetic, and tetragonal orbitally ordered multiferroic phases. (b) Temperature dependence of the cubic (8 0 0) peak. (c) Temperature dependence of the (12 4 0) peak. Both (b) and (c) show a clear splitting at 15 K and below, and a broadening alreadty below 60 K.

Our primary interest is in the onset of tetragonal distortions, i.e., when the difference in the $a$ and $c$ lattice parameters can be resolved. An analysis of the individual peaks in Figs. 2(b) and (c), shows an apparent shoulder at 35 K in the (8 0 0) reflection, and a clear splitting in the (8 0 0) and (12 4 0) reflections, and in all diffraction patterns collected at and below 15 K. Such a resolvable peak splitting diagnostically establishes that the symmetry has been reduced from cubic for these temperatures. Although at temperatures above 15 K a clear splitting cannot be resolved instrumentally, the structural distortion is already manifested in a broadening of these peaks. One possible way to estimate the actual temperature when the material loses cubic symmetry is by extrapolation of the temperature dependency of the low-symmetry lattice parameters – see discussion. Here we only note that we have treated the 35 K data as tetragonal based on: (i) significant peak broadening compared with data at 60 K, and (ii) extrapolated thermal contraction data (see discussion).

**Thermal expansion**

In addition to the synchrotron powder diffraction, we performed thermal expansion experiments to characterize the macroscopic behavior of the sample. Fig. 3 shows the temperature dependence of the thermal expansion coefficient, $\alpha$, calculated from measurements of the change in the sample length on heating shown in the inset of this figure. This dataset has several prominent features: a clear peak marking $T_C \sim 165$ K, a step like change at $T_m \sim 65$ K, and a lambda-type anomaly at $T_{OO} = 9$ K, where $\alpha$ changes the sign from positive to negative.

**Resistivity and polarization**

To understand how the structural properties couple to the electronic orders we measured resistivity and electric polarization on single crystals of the same batch. The data are presented in Figs. 4(a) and (b), respectively. The resistivity data show that the sample is a semiconductor at room temperature ($\rho_{300\,K} \sim 70$ $\Omega$ cm) and becomes highly insulating at low temperature, $\rho_{20\,K} > 10^6$ $\Omega$ cm. There is a hump in the resistivity at $\sim 170$ K, slightly above the onset of ferrimagnetic order at $T_C \sim 165$ K.

The current x-ray experiments provide the resolution of the lattice parameters but powder diffraction cannot resolve the difference between polar and nonpolar space groups. To test for polar phases, we measured electric polarization below 20 K, where the material is insulating enough for such studies. Fig. 4(b) shows the temperature dependence of the polarization, calculated by integration of the pyrocurrent data measured under different poling conditions. Interestingly, there is no anomaly in the polarization data if the sample is cooled after poling in a magnetic field but in the absence of a poling electric field. After poling the crystal in the electric field only, a small polarization $P \sim 2$ $\mu C$ m$^{-2}$ appears below $T_{OO} = 9$ K, whose sign depends on the direction of the applied poling voltage (not shown for clarity). This indicates the sample is ferroelectric, as the sign of the polarization is switchable through the phase transition. Importantly, the polarization at the transition can be significantly increased to $\sim 15$ $\mu C$ m$^{-2}$ with additional presence of a magnetic field, suggesting either that magnetoelectric coupling is important, or that the electric field alone is not enough to orient the magnetostructural domains. Note that the sign of the polarization is independent of the sign of the applied magnetic field, implying an even coupling term. The pyrocurrent curves in the inset of Fig. 4(b) also reveal a second broad feature starting below ~ 6 K, the origin of which is unclear. We cannot rule out that it is related to thermally stimulated discharge currents, which lead to additional polarization of ~ 13 $\mu C\ m^{-2}$, resulting from the integration of pyrocurrent over a wide temperature range below 6 K.

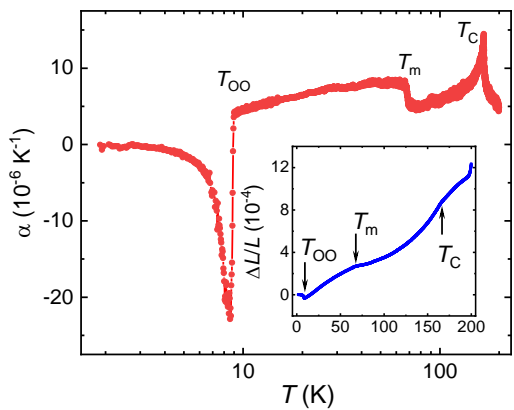

FIG 3: Temperature dependence of the thermal expansion coefficient $\alpha$ of single crystalline FeCr$_2$S$_4$. Inset: temperature dependence of the normalized sample length $\Delta L/L$. Arrows mark the phase transition temperatures.

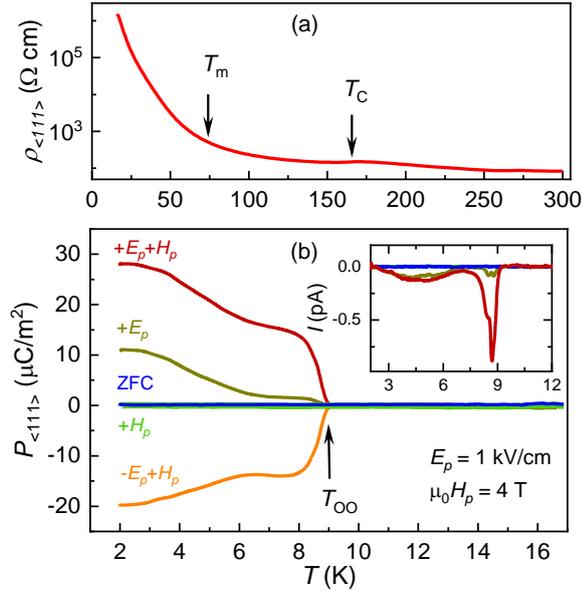

FIG 4: (a) Temperature dependent resistivity measurements. (b) Polarization $P$ vs $T$. The different colors show the different applied fields with the direction of the polarization depending on the sign of the applied electric field, consistent with ferroelectricity. Inset: pyrocurrent $I$ vs $T$ taken on cooling with the fields labelled in (b).

**Magnetization and heat capacity**

Fig. 5 shows the temperature dependences of the magnetization after zero-field cooling (ZFC) and field cooling (FC) as well as that of the specific heat, both measured on the same single crystal. The data agree well with previous results reported in Ref. [14]. We replicate these data to correlate the thermodynamic properties with the structural transitions.

The key observations of the magnetic data are the clear and sudden onset of a finite magnetic moment at $T_C$ ~ 165 K, and a peak at $T_m$ ~ 65 K, where ZFC and FC data diverge, as seen in Fig. 5. In addition, one can observe a small anomaly (increase for FC and decrease for ZFC) in the magnetization at the onset of polarisation at $T_{OO}$ ~ 9 K. In the specific heat $C_p$ there are two clearly pronounced anomalies, one at $T_C$ and another at $T_{OO}$, while no signature of $T_m$ is observed.

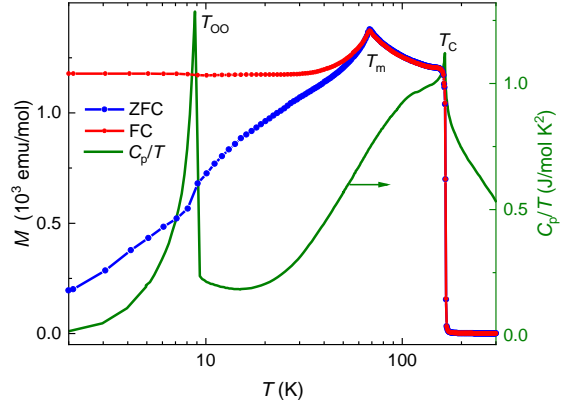

FIG 5: ZFC and FC magnetizations measured in a field of 0.01 T (left axis) and specific heat in representation $C_p/T$ in zero field (right axis) collected on the same sample. Both data sets show a step like anomaly at the onset of the magnetic ordering at $T_C$ ~ 165 K. The magnetic data have an additional peak at $T_m$ ~ 65 K, while a clear peak at $T_{OO}$ ~ 9 K in the specific heat is only slightly pronounced in the magnetization.

## IV. Discussion

For a sake of clarity, we discuss the three transition individually: (i) the high temperature para-to ferrimagnetic transition at $T_C$ ~ 165 K; (ii) the transition at $T_m$ ~ 65 K that is associated with structural distortion and magnetic reorientation, and (iii) further anomalies below 20 K, including the transition at $T_{OO}$.

**Para-to-ferrimagnetic transition**

This transition is clearly manifested in thermodynamic quantities, namely the magnetization and the specific heat (see Fig. 5). There is also a broad peak in the resistivity data in Fig. 4(a), originating from the scattering of the conducting electrons by fluctuations of the magnetic moments that are maximal at $T_C$, giving rise to colossal magnetoresistance [1]. There is no evidence of an anomaly in our x-ray data, with the lattice parameters apparently changing linearly with temperature across this transition, Fig. 2(a). At the same time, the presence of a clear peak in the thermal expansion coefficient at $T_C$, Fig 3, which is associated with exchange striction, suggests that the structural changes do take place although the deviation from the cubic symmetry, which is a necessary consequence of the onset of the long-range ferrimagnetic order, cannot be resolved even in our high-resolution single-crystal x-ray study. For this reason, we refer to the phase between $T_m < T < T_C$ as a pseudo-cubic ferrimagnetic state.

**Cubic-to tetragonal transition**

From our synchrotron data we can conclude that the material is clearly tetragonal at and below 15 K. The precise determination of the onset of the cubic to tetragonal transition is compromised by the resolution of our experiment. While a broadening of specific cubic Bragg peaks is observed below 60 K, their splitting to two peaks is well resolved only at and below 15 K. Fortuitously, when refining the structure below 65 K using the tetragonal space group $I4_1/amd$, the obtained $a$ and $c$ lattice parameters coincide at ~ 65 K and deviate linearly below this temperature as depicted by the yellow dashed lines in Fig. 2(a). This approach is supported and substantiated by the thermal expansion coefficient of Fig. 3, which shows the expected change in trend across $T_m \sim 65$ K. The presence of a structural phase transition can also explain the divergence of the ZFC and FC magnetizations in Fig. 5 and implies a simultaneous change in the magnetic structure. This is naturally expected, since the cubic to tetragonal distortion activates the axial magnetic anisotropy in this compound.

In contrast, neither the resistivity nor the specific heat show a pronounced anomaly at this temperature. This implies that the changes in the phonon energy and the scattering rate of the itinerant electrons are subtle at this magnetostructural transition.

Our synchrotron data indicate that the symmetry of the unit cell changes from $Fd\bar{3}m$ to $I4_1/amd$ through this transition. These structures are schematically depicted in Figs. 6(a) and (b), respectively. Both of these unit cells are viewed down the room temperature $a$-axis. The $a$ and $b$ axes in the low-temperature tetragonal structure are rotated by 45° with respect to the [100] and [010] cubic axes, as schematically shown in Fig. 6(c).

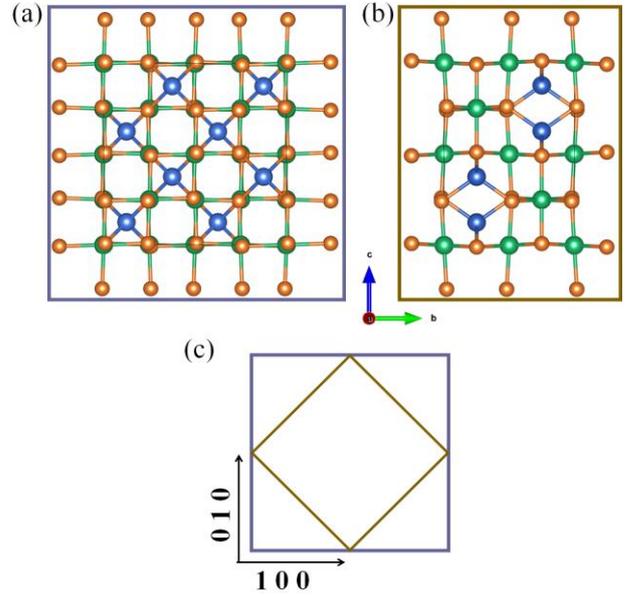

FIG 6: (a) High temperature cubic $Fd\bar{3}m$ unit cell. (b) Low temperature tetragonal $I4_1/amd$ unit cell. In both (a) and (b) the blue atoms are Fe, the green atoms are Cr, and the gold atoms are S. (c) Schematic of the redefined unit cell across the structural transformation. Panels (a) and (b) are made using VESTA [32].

**Onset of polar order**

While powder x-ray diffraction in the present case does not permit the distinction of polar and non-polar tetragonal space groups, the appearance of a peak in pyrocurrent at $T_{OO}$ provides a clear evidence for the onset of a ferroelectric phase as can be followed in Fig. 4(b). This is associated with a step-wise change in the thermal expansion coefficient, and a lambda like anomaly in the specific heat. The fact that the magnitude of the pyrocurrent can be enhanced with a magnetic field, applied in addition to an electric field, strongly suggests this phase is a multiferroic phase.

Curiously, despite the multiferroic nature of this phase, application of magnetic field on its own is not able to establish a net polar moment. In contrast, an electric field leads to finite polarization. Surprisingly, the combination of electric and magnetic field can substantially increase the ferroelectric polarization. This may be understood in the following scenario. Electric field applied along any of the cubic <100> axes, favours and increases the population of the tetragonal structural domain states with *c* axis co-aligned with the field and, at the same time, it also favours the domain state with polarization along the electric field with respect to the domain state with opposite polarization. However, not even the strongest electric fields applied in this study can create a structural mono-domain state and saturate the polarization. The action of the magnetic field, applied along any of the cubic <100> axes, is different: it also favours the tetragonal structural domain states with *c* axis co-aligned with the field, but would not switch their polar moments. For this reason, electric and magnetic fields act cooperatively and yield to increased polarization.

Based on our observations we are able to make some inferences about the symmetry of the ground state. Our synchrotron study confirms that the symmetry in the state below $T_{OO}$ is also tetragonal. While the synchrotron study could not distinguish between the centrosymmetric and polar tetragonal structures, the observation of finite electric polarization below $T_{OO}$ indicates a polar structure of the ground state. The tetragonal magnetic point groups, which allow spontaneous electric polarization are 4, 4′, 4*mm*, 4′*mm*′ and 4*m*′*m*′. Since the present compound exhibits a finite ferromagnetic component, 4 and 4*m*′*m*′ are the most probable magnetic point groups. These are all subgroups of 4/*mmm1*′, which is the point group corresponding to the non-magnetic non-polar space group *I4₁/amd*, refined by our x-ray study.

## V. Conclusions

In this work, we investigated the sequence of the phase transitions within spinel $FeCr_2S_4$ using structural, electronic, thermodynamic, and magnetic measurements. Using high-resolution synchrotron data, we could confirm the previously assumed tetragonal phase at low temperatures, below $T_m \sim 65$ K. Above this temperature, no deviation of the cubic crystal structure was resolved, although the material becomes ferrimagnetic already below $T_C \sim 165$ K. Therefore, we refer to the phase between 65 - 165 K as a pseudo-cubic ferrimagnet. While no further lowering of the symmetry could be traced in our x-ray data down to 4 K, our pyroelectric measurements reveal a transition from the non-polar tetragonal state to a polar tetragonal state at $T_{OO} \sim 9$ K, with the possible magnetic point group symmetries 4 or 4*m*′*m*. As such, we have clarified several long-standing questions in this enigmatic compound, but more work, potentially including direct imaging of the microstructure, is required.

**Acknowledgments**
This work was supported by the Deutsche Forschungsgemeinschaft (DFG) through Transregional Research Collaboration TRR 80 (Augsburg, Munich, and Stuttgart), and by the project ANCD 20.80009.5007.19 (Moldova). DME wishes to thank and acknowledge funding by the DFG individual fellowship, number (EV 305/1-1).